\documentclass[aps,prl,preprint,superscriptaddress]{revtex4}
\usepackage{graphicx}
\usepackage{amssymb}
\usepackage{amsmath}
\usepackage{color}

\begin{document}

\title{Magnetic moment evolution and spin freezing in doped BaFe$_{2}$As$_{2}$}

\author{Jonathan Pelliciari}
\email[]{jonathan.pelliciari@psi.ch}
\affiliation{Research Department of Synchrotron Radiation and Nanotechnology, Paul Scherrer Institut, CH-5232 Villigen PSI, Switzerland}
\author{Yaobo Huang}
\affiliation{Research Department of Synchrotron Radiation and Nanotechnology, Paul Scherrer Institut, CH-5232 Villigen PSI, Switzerland}
\affiliation{Beijing National Lab for Condensed Matter Physics, Institute of Physics, Chinese Academy of Sciences, Beijing 100190, China}
\author{Kenji Ishii}
\affiliation{Synchrotron Radiation Research Center, National Institutes for Quantum and Radiological Science and Technology, Sayo, Hyogo 679-5148, Japan}   
\author{Chenglin Zhang}
\affiliation{Department of Physics and Astronomy, Rice University, Houston, Texas 77005, USA}
\author{Pengcheng Dai}
\affiliation{Department of Physics and Astronomy, Rice University, Houston, Texas 77005, USA}
\author{Gen Fu Chen}
\affiliation{Beijing National Laboratory for Condensed Matter Physics, and Institute of Physics, Chinese Academy of Sciences, Beijing 100190, China}
\author{Lingyi Xing}
\affiliation{Beijing National Lab for Condensed Matter Physics, Institute of Physics, Chinese Academy of Sciences, Beijing 100190, China}
\author{Xiancheng Wang}
\affiliation{Beijing National Lab for Condensed Matter Physics, Institute of Physics, Chinese Academy of Sciences, Beijing 100190, China}
\author{Changqing Jin}
\affiliation{Beijing National Lab for Condensed Matter Physics, Institute of Physics, Chinese Academy of Sciences, Beijing 100190, China}
\affiliation{Collaborative Innovation Center for Quantum Matters, Beijing, China}
\author{Hong Ding}
\affiliation{Beijing National Lab for Condensed Matter Physics, Institute of Physics, Chinese Academy of Sciences, Beijing 100190, China}
\author{Philipp Werner}
\affiliation{Department of Physics, University of Fribourg, Chemin du Mus\'ee 3 CH-1700, Fribourg, Switzerland}
\author{Thorsten Schmitt}
\email[]{thorsten.schmitt@psi.ch}
\affiliation{Research Department of Synchrotron Radiation and Nanotechnology, Paul Scherrer Institut, CH-5232 Villigen PSI, Switzerland}

\date{\today}

\begin{abstract}
Fe-K$_{\beta}$ X-ray emission spectroscopy measurements reveal an asymmetric doping dependence of the magnetic moments $\mu_\text{bare}$ in electron- and hole-doped BaFe$_{2}$As$_{2}$. At low temperature, $\mu_\text{bare}$ is nearly constant in hole-doped samples, whereas it decreases upon electron doping. Increasing temperature substantially enhances $\mu_\text{bare}$ in the hole-doped region, which is naturally explained by the theoretically predicted crossover into a spin-frozen state. Our measurements demonstrate the importance of Hund's coupling and electronic correlations, especially for hole-doped BaFe$_{2}$As$_{2}$, and the inadequacy of a fully localized or fully itinerant description of the 122 family of Fe pnictides.
\end{abstract}

\maketitle
Soon after the discovery of high temperature superconductivity in Fe pnictides \cite{kamihara_iron-based_2008}, antiferromagnetic ordering in the form of a spin density wave has been observed in the parent compounds \cite{stewart_superconductivity_2011, johnston_puzzle_2010}. 
The nature of this antiferromagnetism has been highly debated, as demonstrated by the use of antipodal theoretical descriptions, namely, the itinerant and the localized one \cite{chubukov_itinerant_2015,johnston_puzzle_2010,johnson_iron-based_2015,stewart_superconductivity_2011,dai_antiferromagnetic_2015,dai_magnetism_2012}. In the former, magnetism arises from Fermi surface nesting in a similar way to metallic Cr \cite{fawcett_spin-density-wave_1988}, where this phenomenon leads to spin-density wave ordering due to a diverging susceptibility at the nesting wavevector. In Fe pnictides, the discovery, by means of angle resolved photoemission spectroscopy, of cylindric hole and electron pockets satisfying these nesting conditions supported such a picture, together with the metallic ground state, and apparently low electronic correlations \cite{richard_fe-based_2011, mazin_superconductivity_2010, stewart_superconductivity_2011, johnston_puzzle_2010, graser_near-degeneracy_2009}. However, this weak-coupling scenario could not explain some characteristic properties of Fe pnictides, such as the presence of magnetic moments ($\mu$) at high temperature, outside the antiferromagnetic phase, and the persistence of spin excitations in non-magnetically ordered phases \cite{zhou_persistent_2013, pelliciari_intralayer_2016, gretarsson_spin-state_2013, gretarsson_revealing_2011, simonelli_coexistence_2012, vilmercati_itinerant_2012, bondino_electronic_2010, bondino_evidence_2008, mannella_magnetic_2014, dai_antiferromagnetic_2015}. These two aspects are more consistently explained in a strong-coupling picture, where strong electronic correlations localize the spins as in Mott-Hubbard-like scenarios \cite{yu_mott_2011, si_strong_2008, si_correlation_2009}. However, the metallicity and low $\mu$ of Fe pnictides conflict with such an extreme strong coupling description.

A formalism which can handle both the itinerant and localized nature of electrons is the dynamical mean field theory (DMFT) \cite{georges_1996}. Thanks to fairly recent methodological advances \cite{werner_2006,werner_2007}, this formalism can efficiently handle the strongly-correlated metal regime of multi-orbital Hubbard models, such as those relevant for the description of Fe pnictides. An important theoretical prediction from DMFT studies \cite{werner_spin_2008, liebsch_2010, haule_2009, werner_satellites_2012} is the phenomenon of {\it spin-freezing} (SF). In systems with strong Hund's coupling, long-lived magnetic moments appear in the metal phase, if the filling and interaction strength place the system in the vicinity of the half-filled Mott insulator. 
The magnetic moment has been measured in BaFe$_2$As$_2$ \cite{gretarsson_revealing_2011,johnson_iron-based_2015,johnston_puzzle_2010,stewart_superconductivity_2011,dai_antiferromagnetic_2015}, but scant spectroscopic information is available on the temperature and doping effects on $\mu$. Moreover, the electron itinerancy, i.e. the dynamics of the electrons, leads to quantum fluctuations, which by time-averaging mask the value of $\mu$ observed by slow probes (i.e. neutron diffraction, NMR, and muon relaxation measurements \cite{mannella_magnetic_2014, su_antiferromagnetic_2009,kofu_neutron_2009,kaneko_columnar_2008,zhao_spin_2008,aczel_muon-spin-relaxation_2008, kitagawa_commensurate_2008, kitagawa_antiferromagnetism_2009, pratt_coexistence_2009, laplace_atomic_2009, bonville_incommensurate_2010, lumsden_magnetism_2010, huang_neutron-diffraction_2008, dai_magnetism_2012}), making it difficult to extract the ``bare" value of $\mu$.
Fast spectroscopies, probing at the timescale of the electron dynamics (on the order of femtoseconds), are therefore essential to obtain snapshots of the value of $\mu$. This is achieved by the use of techniques such as photoelectron, X-ray absorption, and X-Ray emission spectroscopy \cite{gretarsson_spin-state_2013, gretarsson_revealing_2011, simonelli_coexistence_2012, vilmercati_itinerant_2012-1, bondino_electronic_2010, bondino_evidence_2008, mannella_magnetic_2014}, which indeed produce higher values of $\mu$ compared to their slower counterparts. Additionally, as explained in Refs.~\cite{vilmercati_itinerant_2012, hansmann_dichotomy_2010}, it is possible to distinguish different aspects of $\mu$, the bare $\mu$ ($\mu_\text{bare}=\langle S_{i}\rangle$) connected to quantum fluctuations and the correlated $\mu$ ($\mu_\text{corr}^2=\langle S_{i}\cdot S_{i+1}\rangle$), which is indicative of dressed quasiparticles (spin excitations). These physical entities represent different aspects of magnetism, have different characteristic time and energy scales, and are probed by different experimental techniques \cite{mannella_magnetic_2014}.
$\mu_\text{bare}$ is detected by local probes such as photoelectron, X-ray absorption, and X-Ray emission spectroscopy \cite{mannella_magnetic_2014,bondino_evidence_2008,simonelli_temperature_2014,gretarsson_spin-state_2013,gretarsson_revealing_2011}, whereas $\mu_\text{corr}$ is measured by employing inelastic spectroscopies, such as inelastic neutron scattering \cite{dai_antiferromagnetic_2015,wang_doping_2013,luo_electron_2013}.

\begin{figure}[t]
\begin{center}
\includegraphics[width=\textwidth]{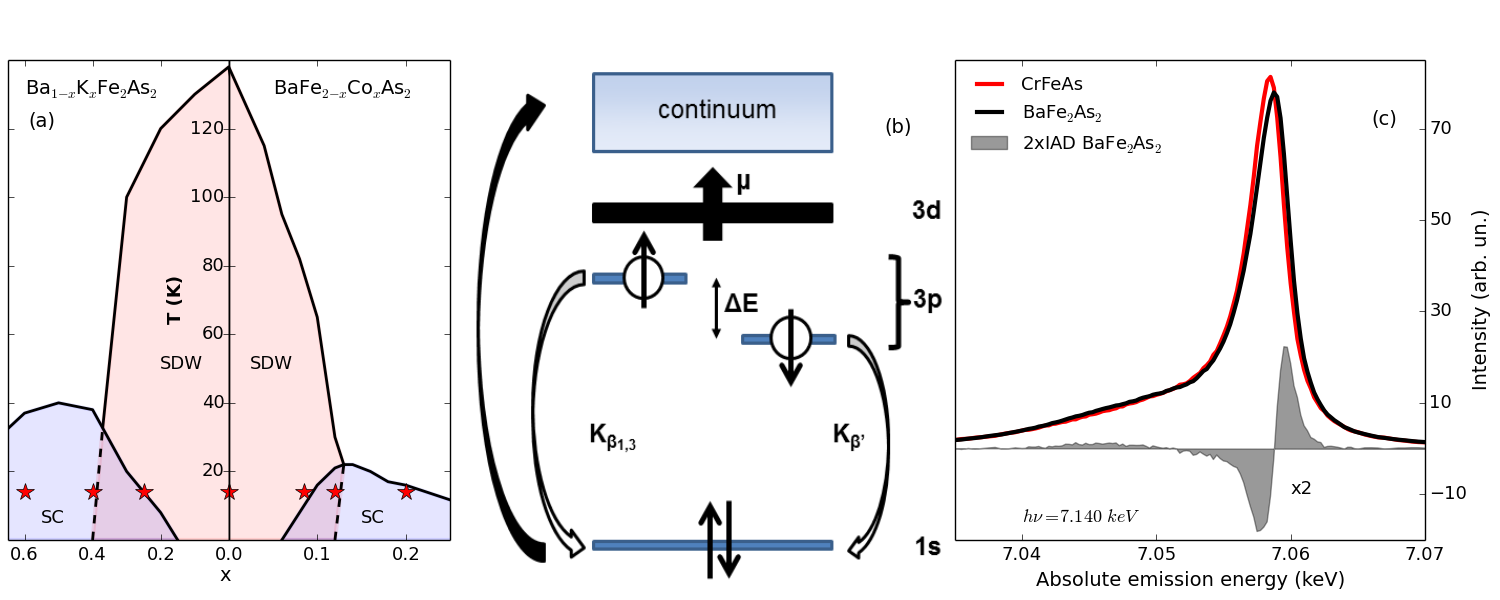}
\caption{\label{fig:fig1} (a) Phase diagram of Ba$_{1-x}$K$_x$Fe$_2$As$_2$ and BaFe$_{2-x}$Co$_x$As$_2$. The red stars depict the doping levels measured. (b) Sketch of the XES process. (c) Exemplary Fe-K$_\beta$ XES for CrFeAs and BaFe$_2$As$_2$ at 15 K. The former is taken as a reference and the IAD is calculated (see main text) and depicted as gray shadowed curve.}
\end{center}
\end{figure}

In this Letter, we present Fe-K$_{\beta}$ X-ray emission spectroscopy (XES) measurements of $\mu_\text{bare}$ in electron- (BaFe$_{2-x}$Co$_{x}$As$_{2}$) and hole-doped (Ba$_{1-x}$K$_{x}$Fe$_{2}$As$_{2}$) Fe pnictides. As outlined in Fig.~\ref{fig:fig1}a by the stars, our study covers a large range of the phase diagram, from underdoped to overdoped for both electron and hole doping. 
As we will show, at 15 K, in hole-doped compounds, $\mu_\text{bare}$ exhibits a weak doping dependence, keeping a value around 1.3 $\mu_B$, typical of the parent compound whereas in electron-doped BaFe$_{2}$As$_{2}$, a decrease is observed, with $\mu_\text{bare}$ being gradually quenched to 1.1 $\mu_B$ for the most overdoped sample. While increasing the temperature to 300~K enhances $\mu_\text{bare}$ in all samples, this effect is more pronounced in hole-doped samples than in electron-doped ones. This shows the inadequacy of a fully itinerant approach to explain the formation of local moments and underlines the importance of Hund's coupling and electronic correlations in Fe pnictides. 
A much more consistent explanation of the doping and temperature evolution of $\mu_\text{bare}$ can be given, with the aid of DMFT calculations, in terms of SF. In BaFe$_{2}$As$_{2}$, the nominal $d^6$ occupation and intermediate strength of the electronic correlations imply that the undoped compound is close to the SF crossover regime \cite{werner_satellites_2012}. Upon hole-doping, as the $d$-filling approaches $n_d=5$ (half-filling), the effect of the Hund's coupling increases, frozen moments appear, and the resulting scattering leads to short quasi-particle life-times and an ill-defined bandstructure. Electron doping, on the other hand, results in a more conventional Fermi-liquid metal, with a well-defined bandstructure and Fermi surface. 
The electronic screening of $\mu$, by a multi-channel Kondo effect \cite{yin_2012}, leads to an unusual temperature dependence: $\mu$ increases with increasing temperature due to a weaker screening effect. Frozen moments with very low Kondo screening temperature appear in the strongly hole-doped region, while  electron doping nudges the system towards a more conventional Fermi liquid state with a reduced $\mu$. 
In the spin-freezing crossover regime, the Kondo screening temperature varies strongly with doping and we hence expect a large temperature variation of the local moment.

XES has been established as an extremely sensitive technique in the detection of $\mu_\text{bare}$ \cite{bergmann_x-ray_2009, vanko_probing_2006, gretarsson_spin-state_2013, gretarsson_revealing_2011, ortenzi_structural_2015, simonelli_coexistence_2012, simonelli_temperature_2014}. In this spectroscopy a core electron from the Fe 1$s$ core shell is excited into the continuum by a photon (in our case $h\nu$=7.140 keV), the core hole is then filled up by a Fe 3$p$ electron together with the emission of a photon ($h\nu$= 7.040 - 7.065 keV), as shown by the scheme in Fig.~\ref{fig:fig1}b. The final state, being Fe 3$p^5$, has a wavefunction partly overlapping with the Fe 3$d$ orbitals, which is consequently affected by the spin polarization of the valence band. This gives rise to a main emission line (composed of K$_{\beta_{1}}$ and K$_{\beta_{3}}$) and a satellite peak (K$_{\beta'}$) as shown in Fig.~\ref{fig:fig1}b. The relative intensity of these peaks directly depends on the Fe 3$d$ net spin \cite{bergmann_x-ray_2009, vanko_probing_2006, gretarsson_spin-state_2013, gretarsson_revealing_2011, ortenzi_structural_2015, simonelli_coexistence_2012, simonelli_temperature_2014}, and employing a calibration procedure, a quantitative determination of $\mu_\text{bare}$ is possible. 
This method probes a fs timescale \cite{mannella_magnetic_2014} allowing the measurement of $\mu_\text{bare} = \langle S_i \rangle$ and minimizing the problem of electron dynamics decreasing the measured value of the moment. 

Single crystals of BaFe$_{2}$As$_{2}$, BaFe$_{2-x}$Co$_{x}$As$_{2}$, and Ba$_{1-x}$K$_{x}$Fe$_{2}$As$_{2}$  have been grown by the flux method as described in Refs.~\cite{wang_peculiar_2009, zhang_neutron_2011}.
We performed XES experiments at BL11XU of SPring-8, Hyogo, Japan. The incoming beam was monochromatized by a Si(111) double-crystal and a Si(400) secondary channel-cut crystal. The energy was calibrated by measuring X-ray absorption of an Fe foil and set to 7.140 keV with $\pi$ polarization. We employed three spherical diced Ge(620) analyzers and a detector in Rowland geometry at ca 2 m distance. The total combined resolution was about 400 meV estimated from FWHM of the elastic line. We scanned the absolute emission energy between 7.02 keV and 7.08 keV and normalized the intensity by the incident flux monitored by an ionization chamber. We carried out measurements at both 15 and 300 K.

In Fig.~\ref{fig:fig1}c, we show XES spectra obtained from CrFeAs and BaFe$_2$As$_2$. The former is employed as a standard material due to $\mu_\text{bare}=0$ on the Fe sublattice, together with a similar Fe coordination to the samples investigated. BaFe$_2$As$_2$ has been employed as the high $\mu_\text{bare}$ standard, setting it to a value of 1.3 $\mu_{B}$ taken from Ref.~\cite{gretarsson_revealing_2011}. To determine $\mu_\text{bare}$, we employed the integrated absolute difference (IAD) method described in Ref.~\cite{vanko_probing_2006}. The areas of the spectra are normalized and the difference to the reference spectrum of CrFeAs is calculated. The integration of this difference gives the IAD, which is proportional to $\mu_\text{bare}$. To calibrate the absolute energy, we aligned in an additional step all the spectra to the center of mass as described in Ref.~\cite{glatzel_high_2005}. We show the IAD obtained for the parent compound as the shadowed part of Fig.~\ref{fig:fig1}c.

\begin{figure}
\includegraphics[scale=0.4]{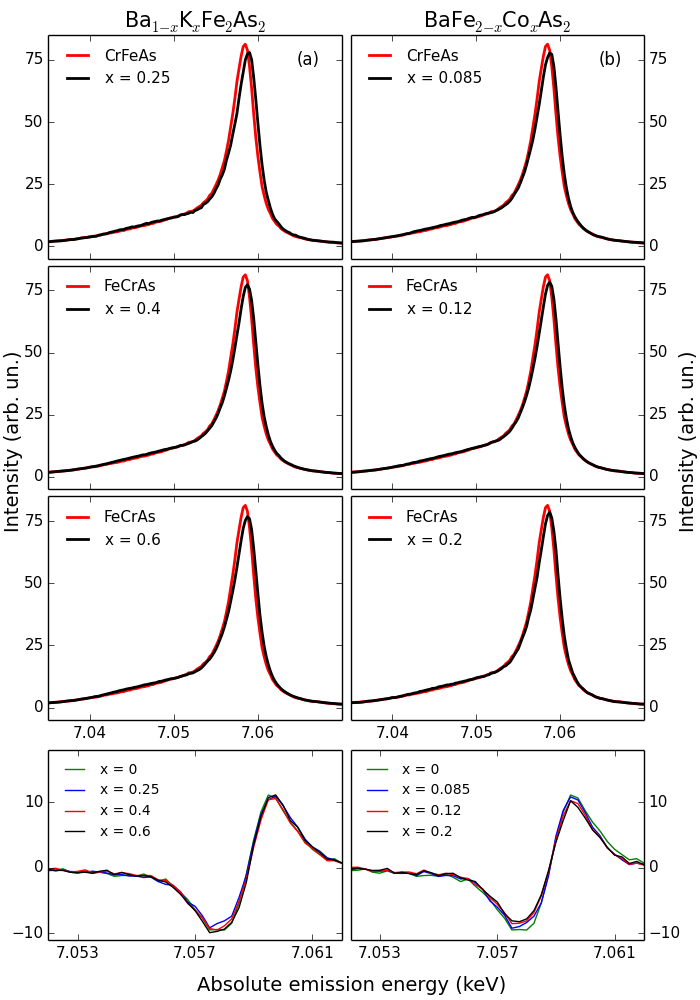}
\caption{\label{fig:fig2} K$_\beta$ XES for Ba$_{1-x}$K$_x$Fe$_2$As$_2$ (a) with $x$ = 0.25, 0.4 and 0.6 and BaFe$_{2-x}$Co$_x$As$_2$ (b) with $x$ = 0.085, 012 and 0.2 at 15 K. The last row is indicating the relative IAD for Ba$_{1-x}$K$_x$Fe$_2$As$_2$ and BaFe$_{2-x}$Co$_x$As$_2$.}
\end{figure}

Having calibrated the instrumental response of IAD vs. $\mu_\text{bare}$, we now quantify $\mu_\text{bare}$ in the doped compounds of BaFe$_2$As$_2$. In Fig.~\ref{fig:fig2}a, we present the evolution of the XES for hole-doped Ba$_{1-x}$K$_{x}$Fe$_{2}$As$_{2}$ samples with $x = 0.25$, 0.4, and 0.6 at 15~K. All spectra look very similar with almost no modification detectable. Consequently, the IAD shown in the bottom panels of Fig.~\ref{fig:fig2}a displays little change of $\mu_\text{bare}$ with hole doping. Moving to the XES spectra of electron-doped BaFe$_{2-x}$Co$_{x}$As$_{2}$ ($x = 0.085$, 0.12, and 0.2) depicted in Fig.~\ref{fig:fig2}b, we observe similar spectral features compared to hole doped BaFe$_2$As$_2$. However, the IAD analysis shows here a decrease of $\mu_\text{bare}$ from 1.3$\pm0.15$ $\mu_B$ to 1.1$\pm0.15$ $\mu_B$ with Co doping. 

\begin{figure}
\includegraphics[scale=0.35]{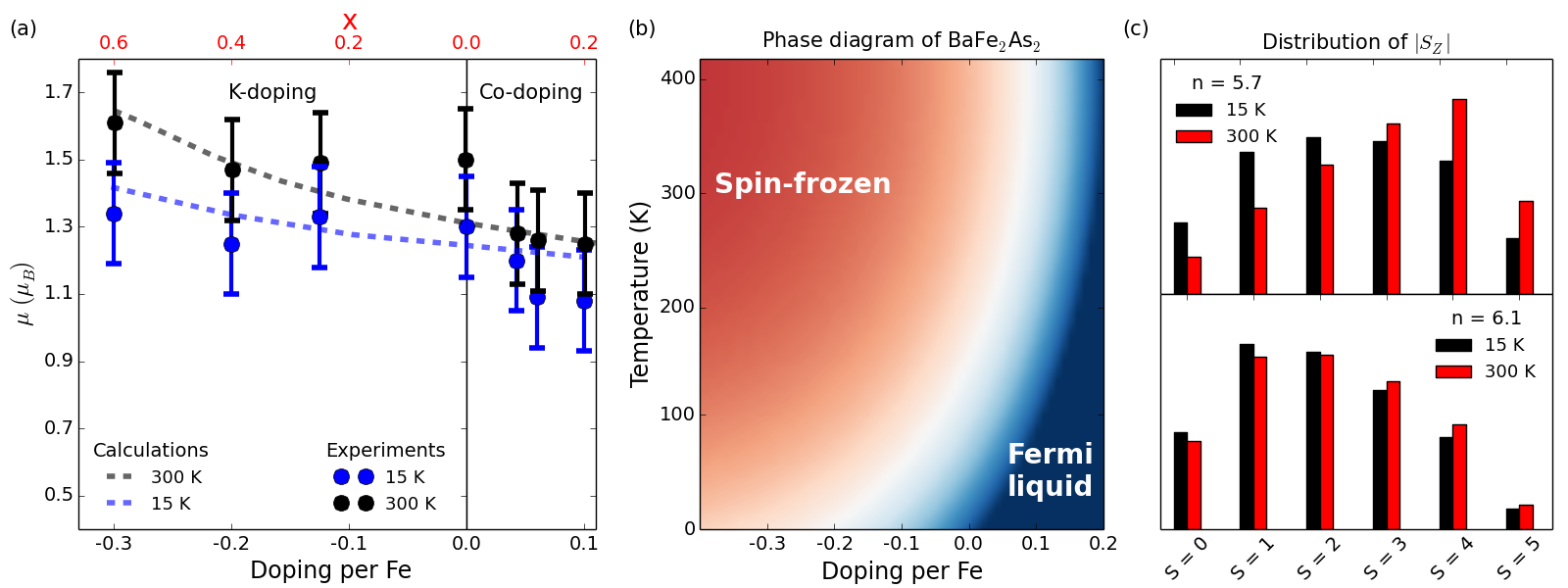}
\caption{\label{fig:fig3} a) Evolution of $\mu_\text{bare}$ for Ba$_{1-x}$K$_x$Fe$_2$As$_2$ and BaFe$_{2-x}$Co$_x$As$_2$. The blue dots with error bars indicate measurements at 15~K, while the black dots with error bars represents $\mu_\text{bare}$ at 300~K. The dashed colored lines are values for $\mu$ obtained from the DMFT calculations. b) Sketch of the theoretical phase diagram for Ba$_{1-x}$K$_x$Fe$_2$As$_2$ and BaFe$_{2-x}$Co$_x$As$_2$ displaying the spin-frozen and Fermi liquid regimes and their evolution with doping and temperature. c) Distribution of $|S_z|$ values in the thermal ensemble for n = 5.7 (top) and n = 6.1 (bottom) at 15 and 300~K.}
\end{figure}

This is summarized in Fig.~\ref{fig:fig3}a, where we plot the extracted $\mu_\text{bare}$ for all dopings. At 15~K, $\mu_\text{bare}$ remains approximately 1.3 $\mu_B$ in the hole-doped compounds and gradually decreases with doping in electron-doped compounds. This variation is remarkable considering the smaller number of electrons doped by Co-doping compared to the holes injected by K-doping as displayed in the bottom scale of Fig.~\ref{fig:fig3}a. At 0.3 doped holes per Fe no change is observed, whereas doping of just 0.1 electrons per Fe induces a 15 \% decrease of $\mu_\text{bare}$.

We can partially explain our observation by considering the fully itinerant limit, where the nesting strength and $\mu$ are connected, and can be quantified by the Lindhard function, which has been observed to evolve asymmetrically upon doping \cite{neupane_electron-hole_2011}. The nesting strength decreases linearly with the injection of electrons, but remains constant up to $x$ = 0.4 for hole-doping where it starts to decrease for even larger doping \cite{neupane_electron-hole_2011}. This could account for the decrease of $\mu_\text{bare}$ upon electron-doping and partially explain the almost constant $\mu_\text{bare}$ for weak hole-doping, but it clearly fails at higher hole doping concentrations.
XES measurements at 300~K furthermore exhibit an {\it increase} of $\mu_\text{bare}$ in all samples compared to the value of $\mu_\text{bare}$ at 15~K (Fig.~\ref{fig:fig3}a and Supplemental Material). 
The lack of magnetic ordering, and the observation of a paramagnetic state with an increased $\mu_\text{bare}$ at 300~K, demonstrates that a Fermi surface nesting scenario completely fails to describe $\mu$ at high temperature. 
Neutron scattering measurements of $\mu_\text{corr}$ show a good agreement with our findings on electron doped samples \cite{luo_electron_2013}, but a decrease is observed on hole doped samples \cite{wang_doping_2013}. Nonetheless, we believe that a comparison of the results of the two techniques is not straightforward and goes beyond the scope of this paper.

To aid the interpretation of the experimental measurements we performed DMFT simulations of a five-orbital Hubbard model with a semi-circular density of states (DOS) of bandwidth 4 eV, which corresponds to the $d$-electron bandwidth of BaKFe$_2$As$_2$ in the local density approximation \cite{werner_satellites_2012}. The Coulomb interaction matrix was taken from Ref.~\cite{werner_satellites_2012}, but rescaled in such a way that the SF crossover in the model with the simplified DOS occurs near $d$-electron filling $n_d=6$ at temperature $T=100$ K. (The fluctuating local moments at the border of the spin-frozen regime lead to a characteristic $\sqrt{\omega}$ frequency dependence of the self-energy \cite{werner_spin_2008}, which can be used to identify this crossover regime.) We solved the DMFT equations using the hybridization-expansion approach \cite{werner_2006}, restricting the solution to paramagnetic metal states. The hybridization-expansion method gives direct access to the fluctuating Fe-$3d$ states, and allows to calculate the instantaneous $\mu$ (here estimated as $\mu\approx \sqrt{\langle S_z \cdot S_z\rangle}$) in the relevant temperature and doping regime.

The calculations yield magnetic moments between 1.25 and 1.65 $\mu_B$, in good agreement with the experimental results. We show the simulation results for temperatures $T=15$~K and 300~K as dashed lines in Fig.~\ref{fig:fig3}a. They display an increase of $\mu$ with hole-doping and a decrease with electron-doping in qualitatively good agreement with the experiments. The doping evolution can be ascribed to a change in the Fe-3$d$ filling, which affects the distribution of atomic states in the thermal ensemble. In particular, electron-doping (hole-doping) moves the system further away from (closer to) filling $n_d=5$, which is needed to realize the maximum spin state in a localized picture. 
(In the experiments, the formal occupation is 3$d^{6.1}$ and 3$d^{5.7}$  at the highest dopings.) Most interestingly, our calculations also predict an increase of $\mu$ with increasing temperature, an effect which is particularly pronounced on the hole-doped side. Within the SF picture, this arises from a weaker Kondo screening of the local moments at high temperature. It is instructive to look at the distribution of $|S_z|$ values in the thermal ensemble, which is plotted in panel (c) of Fig.~\ref{fig:fig3}. Especially on the hole-doped side, these histograms provide clear evidence for a weight shift towards high-spin states and reduced spin fluctuations at the higher temperature.

By correctly reproducing the experimentally observed stronger increase of $\mu_\text{bare}$ with temperature in hole doped samples, our DMFT calculations confirm that this behavior is a signature of a crossover into a spin-frozen state.
Figure~\ref{fig:fig3}b illustrates the consequences of the SF crossover on the nature of the metallic phase. Electron doping leads the system away from the SF crossover region into a more conventional correlated metal regime, indicated by the blue region, where Fermi surface nesting arguments are applicable. On the other hand, hole-doping shifts the Fe configuration towards half-filling, and the strong scattering from frozen moments wipes out the bandstructure and invalidates Fermi surface nesting arguments. This picture is consistent with recent optical measurements showing a non-Fermi liquid response for hole-doped BaFe$_{2}$As$_{2}$ and Fermi liquid behavior for electron-doped BaFe$_{2}$As$_{2}$ \cite{tytarenko_direct_2015}. 

The difference in slopes observed between calculations and experiments in the electron doped region may be explained as a consequence of Fermi surface nesting. On the electron-doped side nesting may prevail over SF, especially at low temperature, since the system is moving away from the SF crossover. This means that the decreases of $\mu_\text{bare}$ is mainly arising from a worsened nesting, an effect which is not captured by DMFT caclulations with a semi-circular DOS. The situation is opposite on the hole-doped side where Hund's coupling and SF effects dominate nesting and $\mu_\text{bare}$ is more strongly affected by local physics. 

In summary, we have measured $\mu_\text{bare}$ in hole- and electron-doped BaFe$_{2}$As$_{2}$ across the phase diagram. At 15~K, we found $\mu_\text{bare}$ to be weakly dependent on hole doping, but to clearly decrease upon electron doping, in agreement with a crossover between a SF phase and a correlated metal phase with well-defined Fermi surface. The asymmetrical increase of $\mu_\text{bare}$ at 300~K results from a competition between electronic screening and Hund's coupling induced local moment formation. The qualitative agreement between the doping and temperature dependence observed in both theory and experiment demonstrates that a SF occurs in hole doped BaFe$_{2}$As$_{2}$, and that both Hund's-coupling and nesting effects are essential for understanding the unconventional metal state of Fe pnictides. 

\begin{acknowledgments}
J.P. and T.S. acknowledge financial support through the Dysenos AG by Kabelwerke Brugg AG Holding, Fachhochschule Nordwestschweiz, and the Paul Scherrer Institut. The synchrotron radiation experiments were performed at BL11XU of SPring-8 with the approval of the Japan Synchrotron Radiation Research Institute (JASRI) (Proposals No. 2014A3502 and 2014B3502). We thank Y. Shimizu for the support during the experiments at SPring-8 and D. Casa for fabrication of the Ge(620) analyzers. The DMFT calculations were run on the Brutus cluster at ETH Zurich. This research was partly supported by the NCCR MARVEL, funded by the Swiss National Science Foundation. The single crystal growth work at Rice is supported by the US DOE, BES under Contract No. DE-SC0012311 (P.D.). Part of the materials work at Rice is also supported by the Robert A. Welch foundation Grant No. C-1893 (P.D.).
\end{acknowledgments}

\end{document}